\documentstyle{mn}
 
\newif\ifAMStwofonts
\AMStwofontstrue
 
\def\gsimeq{{_>\atop^{\sim}}}
\def\lsimeq
{\hbox{\raise0.5ex\hbox{$<\lower1.06ex\hbox{$\kern-1.07em{\sim}$}$}}}
\def\ont#1{\hfill #1 \hfill}

\ifoldfss
  \ifCUPmtlplainloaded \else
    \NewTextAlphabet{textbfit} {cmbxti10} {}
    \NewTextAlphabet{textbfss} {cmssbx10} {}
    \NewMathAlphabet{mathbfit} {cmbxti10} {} % for math mode
    \NewMathAlphabet{mathbfss} {cmssbx10} {} %  "   "    "
  \fi
  \ifAMStwofonts
    \ifCUPmtlplainloaded \else
      \NewSymbolFont{upmath} {eurm10}
      \NewSymbolFont{AMSa} {msam10}
      \NewMathSymbol{\upi}     {0}{upmath}{19}
      \NewMathSymbol{\umu}     {0}{upmath}{16}
      \NewMathSymbol{\upartial}{0}{upmath}{40}
      \NewMathSymbol{\leqslant}{3}{AMSa}{36}
      \NewMathSymbol{\geqslant}{3}{AMSa}{3E}

      \let\leq=\leqslant 
      \let\geq=\geqslant 
    \fi
  \fi
\fi % End of OFSS
 
\ifnfssone
  \newmathalphabet{\mathit}
  \addtoversion{normal}{\mathit}{cmr}{m}{it}
  \addtoversion{bold}{\mathit}{cmr}{bx}{it}
  \newmathalphabet{\mathbfit} % math mode version of \textbfit{..}
  \addtoversion{normal}{\mathbfit}{cmr}{bx}{it}
  \addtoversion{bold}{\mathbfit}{cmr}{bx}{it}
  \newmathalphabet{\mathbfss} % math mode version of \textbfss{..}
  \addtoversion{normal}{\mathbfss}{cmss}{bx}{n}
  \addtoversion{bold}{\mathbfss}{cmss}{bx}{n}
  \ifAMStwofonts
    \ifCUPmtlplainloaded \else
      \UseAMStwoboldmath
      \makeatletter
      \new@mathgroup\upmath@group
      \define@mathgroup\mv@normal\upmath@group{eur}{m}{n}
      \define@mathgroup\mv@bold\upmath@group{eur}{b}{n}
      \edef\UPM{\hexnumber\upmath@group}
      \new@mathgroup\amsa@group
      \define@mathgroup\mv@normal\amsa@group{msa}{m}{n}
      \define@mathgroup\mv@bold\amsa@group{msa}{m}{n}
      \edef\AMSa{\hexnumber\amsa@group}
      \makeatother
      \mathchardef\upi="0\UPM19
      \mathchardef\umu="0\UPM16
      \mathchardef\upartial="0\UPM40
      \mathchardef\leqslant="3\AMSa36
      \mathchardef\geqslant="3\AMSa3E

      \let\leq=\leqslant 
      \let\geq=\geqslant 
    \fi
  \fi
\fi % End of NFSS release 1
 
\ifnfsstwo
  \DeclareMathAlphabet{\mathbfit}{OT1}{cmr}{bx}{it}
  \SetMathAlphabet\mathbfit{bold}{OT1}{cmr}{bx}{it}
  \DeclareMathAlphabet{\mathbfss}{OT1}{cmss}{bx}{n}
  \SetMathAlphabet\mathbfss{bold}{OT1}{cmss}{bx}{n}
  \ifAMStwofonts
    \ifCUPmtlplainloaded \else
      \DeclareSymbolFont{UPM}{U}{eur}{m}{n}
      \SetSymbolFont{UPM}{bold}{U}{eur}{b}{n}
      \DeclareSymbolFont{AMSa}{U}{msa}{m}{n}
      \DeclareMathSymbol{\upi}{0}{UPM}{"19}
      \DeclareMathSymbol{\umu}{0}{UPM}{"16}
      \DeclareMathSymbol{\upartial}{0}{UPM}{"40}
      \DeclareMathSymbol{\leqslant}{3}{AMSa}{"36}
      \DeclareMathSymbol{\geqslant}{3}{AMSa}{"3E}

      \let\leq=\leqslant 
      \let\geq=\geqslant 
    \fi
  \fi
\fi % End of NFSS release 2
 
\ifCUPmtlplainloaded \else
  \ifAMStwofonts \else % If no AMS fonts
    \def\upi{\pi}
    \def\umu{\mu}
    \def\upartial{\partial}
  \fi
\fi

 \title[Radio Observations of the Marano Field]
       {Radio Observations of the Marano Field and the Faint Radio Galaxy
         Population}
 \author[C. Gruppioni et al.]
        {C.~Gruppioni,$^{1,2}$\thanks{Present Address: Astrophysics Group,
        Imperial College London, Blackett Laboratory, Prince Consort Road, 
        London SW7 2BZ.}
        G.~Zamorani,$^{3,2}$ H.R.~de Ruiter,$^{3,2}$ P.~Parma,$^2$ 
\newauthor
        M.~Mignoli,$^3$ and C.~Lari,$^2$\\
 $^1$Dipartimento di Astronomia, Universit\`a di Bologna, via Zamboni 33,
     I--40126 Bologna, Italy\\
 $^2$Istituto di Radioastronomia del CNR, via Gobetti 101, I--40129 Bologna,
     Italy\\
 $^3$Osservatorio Astronomico di Bologna, via Zamboni 33, I--40126 Bologna,
     Italy} 
\date{Accepted 1996 November 21. Received 1996 July 26}
 \pagerange{\pageref{firstpage}--\pageref{lastpage}}
\pubyear{1996}

\begin{document}
 
\label{firstpage}
 
\maketitle

\def\gsimeq{{_>\atop^{\sim}}}
\def\lsimeq
{\hbox{\raise0.5ex\hbox{$<\lower1.06ex\hbox{$\kern-1.07em{\sim}$}$}}}
\def\ont#1{\hfill #1 \hfill}

\begin {abstract}
%\baselineskip = 0.6truecm

Radio surveys with the Australia Telescope Compact Array 
have been carried out at 1.4 and 2.4 GHz with a limiting
flux of $\sim$0.2 mJy at each frequency in the {\it Marano
Field}, in which deep optical and X-ray (ROSAT) data are also available. In
this paper we present the
two radio samples, complete at the 5$\sigma_{local}$ level,
consisting of 63 and 48 sources respectively at 1.4 and 2.4 GHz.
The 1.4 GHz normalized differential source counts show a flattening below 
about one mJy, in agreement with the results from previous surveys.
The 2.4 GHz counts, which are the deepest at this or
similar frequencies (e.g. 2.7 GHz), agree well with the 2.7 GHz
counts at higher fluxes and with the extrapolations down to $\sim$2 mJy
based on fluctuation analyses. 

The 2.4--1.4 GHz spectral index distributions are presented for both the
complete samples in two flux density intervals. The median spectral index
for the 1.4 GHz sample remains constant at $\alpha \sim$0.8, down to the lowest
fluxes ($S_{1.4} \simeq 0.2$ mJy), while for the higher frequency sample
the spectral index distribution flattens in the lower flux 
density interval ($S_{2.4} < 0.8$ mJy). 
A significant number of sources with inverted spectrum ($\alpha < 0$) 
does appear in both samples, at low flux level ($\lsimeq$ 2 mJy). These
sources, which are about 25\% of the complete sample at 2.4 GHz,
are probably the ``bright'' counterpart of the inverted spectrum sources
which appear to be almost 50\% of the sources in the even deeper
radio surveys (at $\sim$20--40 $\mu$Jy).

\end{abstract}

\vfill
\eject
 \section {Introduction}

Deep multi--frequency radio source counts, 
together with optical photometric
and spectroscopic identifications, are necessary to
understand the properties and the cosmological evolution of the faint radio 
source populations. 
Deep 1.4 GHz counts show a change in slope below a few milliJansky (mJy), 
corresponding to a more rapid increase in the number of faint sources 
(Windhorst 1984; Windhorst et al. 1985). This change in slope has been
confirmed by independent surveys of the same areas (Oort \& Windhorst
1985) and of different areas (Condon \& Mitchell 1984; Oort 1987).
A similar change in slope below a few mJy is visible also 
at other frequencies, e.g. at 5 GHz (Kellerman et al.
1986; Fomalont et al. 1991) and at 8.44 GHz (Windhorst et al. 1993).
It is well known that the fraction of sources identified with
elliptical galaxies and quasars,
which are the dominant populations in bright radio samples, decreases
significantly at these faint fluxes. 
Various models for different classes of objects, with or without evolution,
have been developed in the recent
literature to explain the observed sub-mJy counts. Evolving models
include normal spiral galaxies (Condon 1984, 1989) or actively star-forming
galaxies (Windhorst 1984; Kron, Koo \& Windhorst 1985; Windhorst et al. 1985,
1987; Oort 1987; Rowan--Robinson et al. 1993). Wall et al. (1986) and
Subrahmanya \& Kapahi (1983), instead, proposed a non-evolving population of 
local (z $<$ 0.1) low-luminosity radio galaxies as an explanation for the
sub-mJy counts. 

Optical identifications for a few samples show that $\sim$(20--30)\% of the
radio sources in the sub-mJy regime are identified with optical 
counterparts brighter than  $m_B \leq$ 22.0, while $\sim$50\% are identified
in deeper CCD frames reaching  $m_B\sim$24.0--25.0. These analyses 
show that the majority of the identified radio sources at the sub-mJy level 
are faint blue galaxies (Kron, Koo \& Windhorst 1985; Windhorst et al.
1985, 1987, 1995; Thuan \& Condon 1987),
often showing peculiar optical morphology indicative of 
interacting, starburst or merging galaxies (Kron, Koo \& Windhorst 1985;
Windhorst, Dressler \& Koo 1987) and seem to occur preferentially in pairs or
small groups (Windhorst et al. 1995). 
This result is confirmed by the
largest spectroscopic work so far available (Benn et al. 1993), which has proved
that most of the optical counterparts with  $m_B \lsimeq$ 22.0 
of the sub-mJy sources
have spectra similar to those of star--forming IRAS galaxies. Note, however,
that the Benn et al. sample of spectroscopic identifications, although
relatively large (58 redshifts), corresponds to 
%GZ
a small fraction (slightly more than 10\%) of the total number of radio 
sources in their sample.
%GZ

The aim of this paper is to investigate the weak radio source population
in an area of the sky, the {\it Marano Field}, which 
has been deeply surveyed
in the optical and X-ray (ROSAT) bands.
In this field a deep ROSAT observation ($\sim$60 ksec) has recently been 
carried out. About fifty X-ray sources have been detected
in the inner 15$^{\prime}$ radius circle at a flux limit of 
$\sim$4$\times$10$^{-15}$ erg cm$^{-2}$ s$^{-1}$ (Zamorani et al. in
preparation). This inner part of the ROSAT
field has been almost entirely covered by CCD exposures, in the
U, B, V and R bands, taken with the ESO NTT, while a larger area ($\sim$1$^o$
diameter) has been covered by ESO 3.6-m plates in the U, J and F bands.
These optical data have been used to select a sample of $\sim$70
spectroscopically confirmed quasars with $J \leq$ 22.5. Fifty--two of these 
quasars constitute a complete sample with $J \leq$ 22.0 
(the MZZ sample, see Zitelli et al. 1992). The radio properties of this sample
of quasars will be discussed in a future paper (Gruppioni et al., in 
preparation).

We observed this field at 1.4 and 2.4 GHz
with the Australia Telescope Compact Array (ATCA). 
In this paper we present two samples
of sources, complete at the 5$\sigma_{local}$ level, corresponding to 
$\sim$0.2 mJy  at each frequency, extracted
from an area of $\sim$0.36 sq. deg. Our goals are (a) to measure the
1.4 and 2.4 GHz counts down to $\sim$0.2 mJy and to compare them
with the existing counts at similar frequencies (Windhorst et al. 1985,
1987 at 1.4 GHz; Condon 1984; Wall \& Peacock 1985 at 2.7 GHz);
(b) to determine the two point
spectral index distribution of these sources and
to study it for different flux density
intervals for both the 1.4 and 2.4 GHz selected samples.

Radio observations, data reduction and the radio catalogues are
described in \S 2. In \S 3 we present the 1.4 and 2.4 GHz
counts.
In \S 4 we discuss the spectral index distribution for different flux
density intervals. 
Our conclusions are given in \S 5.

\vspace{1cm} 

  \section {The Radio Observations}

  \subsection {The ATCA Observations}
 
The radio observations were carried out on 1994 January 4, 5, 6 and 7.
They were made with the Australia Telescope Compact Array (ATCA) 
simultaneously at two different frequencies: 1.380 and 2.378 GHz (referred to
as 1.4 and 2.4 GHz in the rest of the paper).
In order to obtain a good coverage of the inner region of the ROSAT 
field ($\sim$20$^{\prime}$
radius) within the FWHM of the primary beam (33$^{\prime}$ at 1.4 GHz
and 22$^{\prime}$ at 2.4 GHz),
a mosaicing pattern of four separate observations with different pointing 
positions was utilized.
The four observations (12 hours each) were pointed at the vertices
of a square with a side of 16 arcmin centered at the position
of the Marano Field center ($\alpha$(2000) = 03$^h$ 15$^m$ 09$^s$, 
$\delta$(2000) = $-55^o$ 13$^{\prime}$ 57$^{\prime \prime}$).
The observing bandwidth was 128 MHz, consisting of 32$\times$4 MHz channels.
The primary flux density calibrator was PKS B1934$-$638, which was assumed to
have flux densities of 16.24 and 13.05 Jy respectively at 1.4 and 2.4 GHz.
As a phase and secondary amplitude calibrator the source PKS B0302$-$623 was
used.

  \subsection {Data Reduction}

The data were calibrated and reduced using the ATCA reduction package
MIRIAD (Multi--channel Image Reconstruction Image Analysis and Display).
First, each pointing was calibrated and reduced separately as a standard
12 hour exposure. After flagging of bad data and calibrating by using
primary and secondary calibrators, the data were imaged utilizing the
{\it multi--frequency synthesis} procedure available in MIRIAD. 
This procedure creates a single continuum
image from a variety of frequencies. 
In this way
the individual channels are gridded with their correct location in the
{\it u--v} plane, rather than with some average location. Because of our large
observing bandwidth (128 MHz), the {\it multi--frequency synthesis} technique
was necessary, since it allowed us to reduce the bandwidth smearing and
to obtain a better {\it u--v} coverage in a better beam.
For each of the four single fields a 2048$\times$2048 pixel image
was constructed, with a pixel--size of 2.2$\times$2.2 arcsec.

%
%{\it A serious problem which affected our data was an interference which looked
%like a DC--offset at 1.4 GHz.
%This gave rise to significant artefacts particularly near the center of 
%each single pointing map. We
%were not able to find a solution for this problem by using the standard
%procedure usually adopted to correct the correlator DC--offsets.
%By analyzing the data in all baselines and in all channels, we found that
%it was possible to significantly reduce the interference
%by restricting the usable band. At the end we decided to use data only from
%channel 12 to channel 27.} 
%
%@@TO REDUCE AN INTERFERENCE WHICH LOOKED LIKE A DC--OFFSET AND AFFECTED OUR
%1.4 GHZ DATA, WE WERE FORCED TO RESTRICT THE USABLE BAND, BY ELIMINATING THE
%CHANNELS CONTAINING THE INTERFERENCE SIGNAL. THUS,@@
%since the sensitivity is inversely proportional to the
%bandwidth square--root, by restricting the band we increased the limiting
%
%GZ
In order to reduce an interference which looked like a DC--offset and
affected significantly our 1.4 GHz data, we were forced to restrict the usable
band, using data only from channel 12 to channel 27,
%GZ
%In order to reduce an interference present in the observations at 1.4 GHz, we
%restricted the number of channels in the usable band, 
thus increasing the limiting flux at this frequency by about 40\%. 
Moreover, the bandwidth  restriction caused the
reference frequency to be 1.370 GHz instead of the nominal 1.380 GHz.  

For each frequency the four dirty maps were CLEANed separately. 
The source components were then restored 
by convolving them with a beam of
14$^{\prime \prime}$.5$\times$9$^{\prime \prime}$.5 (FWHM), p.a. 15$^o$
and 9$^{\prime \prime}$.8$\times$5$^{\prime \prime}$.0, p.a. 13$^o$
respectively for the 1.4 and 2.4 GHz maps. The final maps at the two frequencies
were then obtained by combining the four single maps
using the mosaic procedure available in MIRIAD.

The minimum rms noise obtained in the central area of the field and in
areas far from bright sources was comparable to the expected one, but
the final noise resulted somewhat structured and irregularly
distributed at low signal--to--noise ratio levels,
%GZ
especially in the 1.4 GHz map. 
%GZ
In order to select a sample 
above a given threshold, defined in terms of {\it local} signal--to--noise ratio,
a detailed analysis of the spatial rms noise distribution in the images was
necessary.
%
%, as described in the next sub--Section.
%
%  \subsection {The Rms Noise Distribution Analysis}
%
%First, we tried to solve both the source extraction problem and the estimate
%of the local rms noise using the task SAD, recently implemented
%in the software package AIPS. SAD is a procedure which is meant to find 
%all the sources above a given threshold, to fit bidimensional Gaussian 
%components to these sources and to create a map of
%residuals by subtracting the fitted sources. 
%
%
%{\it In order to optimize the source extraction, we used multiple runs of SAD by
%decreasing at each run the flux threshold
%on the residual images. Despite this procedure,
%we found that the final results obtained with SAD were not
%fully reliable.} 
%
%
%But, at low flux levels the number of apparently spurious sources found by SAD
%was too large ($\sim$20 \% above the 5.5$\sigma_{local}$ level at 
%1.4 GHz). Moreover, in some cases the flux measured for the faintest
%sources resulted to be significantly overestimated and this is probably the
%cause of the large number of spurious sources.

We were then forced to construct and study in detail
the rms noise map. To do this we utilized the NOAO reduction package
IRAF (Image Reduction and Analysis Facility). The adopted procedure
started with the creation of a {\it background} image,
obtained by fitting every line of the map with a continuous function and 
clipping out every pixel with an absolute value greater
than a fixed number of times the median value.
The clipped pixels were substituted with their corresponding median value.
We then computed a running 
mean of the {\it background} map and subtracted this mean from the 
{\it background} to obtain the residual image,
from which the pixel by pixel rms noise map was finally derived.

%
%{\it The rms noise maps for both frequencies in the central area of 
%36$^{\prime} \times 36^{\prime}$ are shown as grey scale plots in figures 
%1a and 1b. 
%
%In fig. 1a (1.4 GHz) the spatial irregularity of the noise distribution is clearly 
%noticeable, particularly in correspondence of the strongest sources, where
%a sort of correlated rms noise structure does form. The high rms
%noise area at the top left corner of the map is due to the presence of a
%strong radio source ($\sim$0.5 Jy) outside the $36^{\prime} \times 36^{\prime}$
%area. Fig. 1b (2.4 GHz) shows a more regular rms noise distribution, although 
%also in this case higher rms noise values are present in
%correspondence of the strongest sources. Note that, because of the smaller size
%of the 2.4 GHz primary beam, at this frequency
%there are four separated minima of noise corresponding to the centers of each 
%single field, instead of a single minimum in the mosaic center as in the 1.4 
%GHz map.
%
%The area showed in fig 1a--1b is the one}
% 
%
The area from which we have extracted the complete samples of sources described
in the next sub--Section is $36^{\prime} \times 36^{\prime}$.
Outside this area the beam attenuation
increases significantly the limiting flux at both frequencies. 

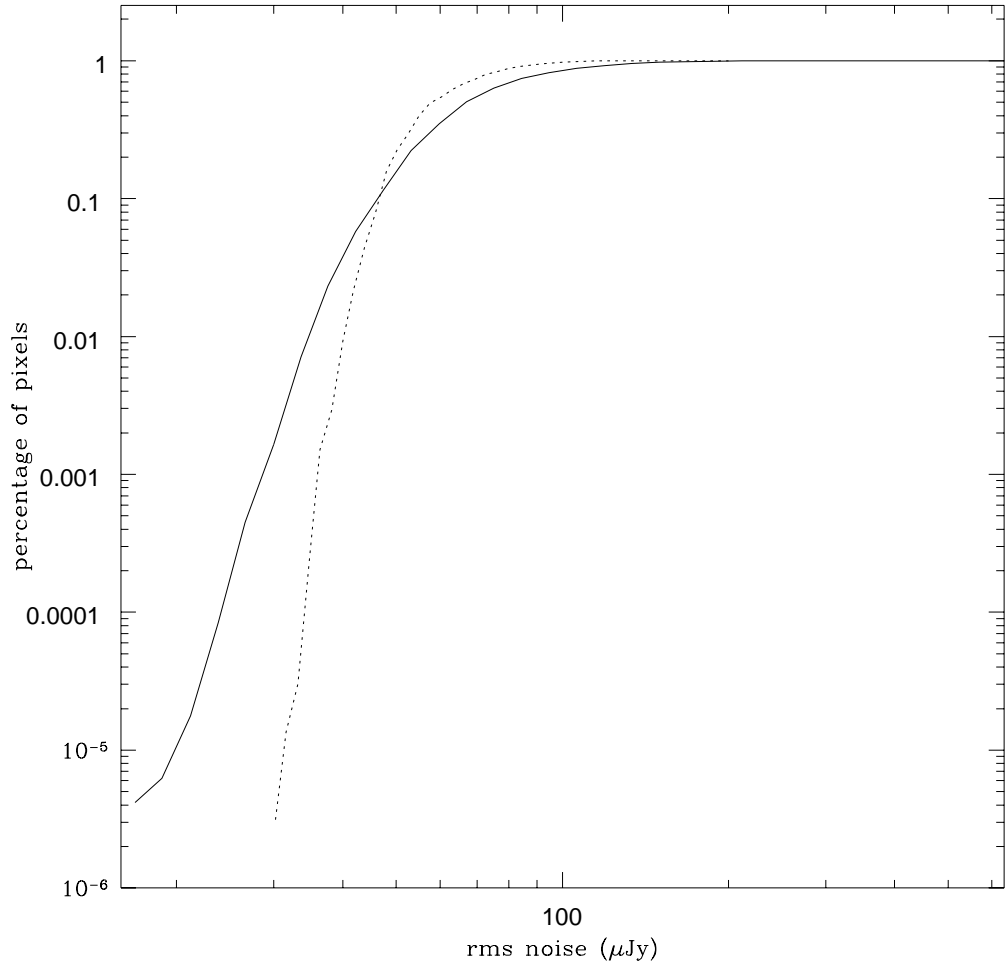
Figure 1 shows the integral distributions of the percentage of pixels with a
given rms noise
as a function of rms noise for the two maps. The steeper curve for the 2.4 GHz
map corresponds to the fact that the noise is significantly more uniform
at this frequency: 
95\% of the pixels have an rms value in the range 40--150 $\mu$Jy at
1.4 GHz and in the range 45--100 $\mu$Jy at 2.4 GHz. These rms integral
distributions have been used to compute the {\it visibility areas} as a 
function of flux which are necessary in order to properly derive the source 
counts (see Section 3).

  \subsection {The 1.4 GHz and 2.4 GHz Samples}

%GZ
At both frequencies we decided to select all sources whose peak flux density 
is greater than or equal to 5 times the local noise. Operatively,
this was done by dividing the original map by the rms noise map described
in Section 2.2 above 
and considering as real sources all the objects showing a signal--to--noise 
ratio greater than or equal to 5. 
%GZ
%Having constructed the rms noise images, the source selection was made by
%simply dividing the original map 5 times by the rms noise map 
%and considering as real sources all the objects showing a signal--to--5
%noise ratio greater than or equal to 1.  We required,
%in fact, that the peak flux density of each source be greater than or equal
%to 5 times the local noise. 

The 36$^{\prime} \times 36^{\prime}$ area
corresponds to about 4.5 $\times 10^{4}$ and 1.2 $\times 10^{5}$ beams at
1.4 and 2.4 GHz respectively. If the noise had been well behaved and gaussian
distributed we would have expected less than one spurious source above
5$\sigma_{local}$ at each frequency. Since this is not the case, in order to 
have an estimate of the number of the possibly spurious sources we checked for
the presence of negative peaks with absolute
value greater than 5$\sigma_{local}$. Indeed, a few such peaks are present
in the two maps, but all of them 
are found in the vicinity of the strongest and extended radio sources,
probably due to unusually deep negative sidelobes, which are difficult to 
clean completely. Thus we are confident that 
at most one or two spurious sources are present in the samples.
%
%Contour representations for the signal--to--noise
%images at the two frequencies are shown in figures 3a and 3b, in which only
%the sources above the 5$\sigma_{local}$ threshold are plotted. 
%Each source
%is labeled with a number corresponding to the numbers given in Tables 1 and
%2 (see below).

%GZ
The source parameters were derived by least--square
fitting of the source surface brightness distribution using an extended 
two--dimensional Gaussian. The parameters derived by the fit are the source
position, the peak and total fluxes ($S_p$ and $S_t$), the Gaussian
half--width and the
position angle (deconvolved with the half--power width of the beam).
For a few faint sources, by comparing the values of each pixel in the data
with the corresponding best fitted values, we found that the Gaussian 
fitting algorithm  produced a significant overestimate of the peak flux
(see Condon 1996 for an extensive discussion about errors in Gaussian
fits). For these sources we derived the peak flux by a second degree
interpolation and the total flux by integrating the map values in a
rectangle around the sources. 
%GZ
%For a few faint sources, for which the Gaussian fitting algorithm 
%produced a significant overestimate of the peak flux,
%we derived the peak flux by a second degree interpolation 
%and the total flux by integrating the map values in a
%rectangle around the sources. 
%We could derive that the peak flux was overestimated because the fitting
%algorithm shows the values of each pixel in the map before fitting, the 
%residual
%values of fitting and the best fitted values. In that way it is possible to
%make direct comparisons between the real peak value and the fitted one.
%See also Condon (1996) for an extensive discussion about errors in Gaussian
%fits.
 
Tables 1 and 2 contain the data for the complete samples at 1.4 and 2.4 GHz,
respectively. The Tables are arranged as follows:

{\it Column (1)} gives the source number (in right ascension order).
For double or multiple sources (clearly resolved into two or
more components) the components are
labeled "A", "B", etc., followed by a line in which parameters for
the total source are given. In a few cases single components in multiple
sources below the completeness
limit of the samples are also listed and are marked with an asterisk.
{\it Columns (2) and (3)} give the right ascension and declination 
of each single component for equinox 2000. For
multiple sources the positions have been computed as the flux weighted
average position for all the components. 
{\it Columns (4) and (5)} give the peak and total flux density,
each with its error (in mJy).
{\it Columns (6) and (7)} give the deconvolved largest angular size 
(in arcsec) and the source position angle (in degrees) for
resolved sources. 
{\it Column (8)} gives the peak signal--to--noise ratio. 

%GZ
By comparing the positions and the position angles of the sources
detected at both frequencies we found that the average error
in both right ascension and declination is $\sim$0.7 arcsec and the
average error in position angle is $\sim$15 degrees. Both
these errors are significantly larger than the formal errors produced
by the fit. For this reason the formal errors for these parameters
are not given in the Tables. In addition
to these statistical errors, there appears to be a systematic difference 
of $\sim$0.5 arcsec in right ascension between the 1.4 and 2.4 GHz maps,
the origin of which is not understood.
%GZ
%We do not report in the table the formal errors produced by the fit for
%the positions and the position angles because they appear to be significantly
%underestimated with respect to the difference in position and position
%angle found between 1.4 and 2.4 GHz for the sources common to both samples.
%In fact, we decided to estimate average errors for these
%quantities by comparing the values obtained for the same sources at the two
%frequencies. From this comparison we conclude that the average error
%in both right ascension and declination is $\sim$0.8 arcsec and the
%average error in position angle is $\sim$15 degrees. In addition to these
%statistical errors, there appears to be a systematic difference of $\sim$
%0.5 arcsec in right ascension between the 1.4 and 2.4 GHz maps, the origin of
%which is not understood.

The 1.4 and 2.4 GHz complete samples contain 63 and 48 sources respectively.
Forty-three sources are in common to the two samples; 9 of
the 20 sources which are part of the 1.4 GHz complete sample only are
detected at 2.4 GHz above 3$\sigma_{local}$, while 3 of the 5 sources which are
part of the 2.4 GHz complete sample only are detected at 1.4 GHz at the same
$\sigma_{local}$ level.
 
%As clearly seen in figures 3a and 3b, 
There are a few borderline cases
in which it is not easy to distinguish between a close pair of unrelated
sources and one source composed of several components. 
In Tables 1 and 2 there are six sources which we have considered double or 
multiple in either the 1.4 GHz or in the 2.4 GHz map 
or in both. Three of these sources have an angular distance (d) between 
different components smaller than 15 arcsec, while the other three have 
15 $<$ d $<$ 25 arcsec. 
The number of expected random pairs of unrelated sources with d $\lsimeq$
25 arcsec, computed on the basis of the observed surface density in our fields, is
about 1 or 2.
%On the basis of the surface density of sources
%in our field {\bf @@WHOSE $\theta(s)$ LAW WAS USED??@@} we expect about 1 or 2 
%pairs of unrelated sources with d $<$ 25 arcsec. 
%
Actually, there are two other pairs of 5$\sigma$ components 
which are at about 25 arcsec from each other, namely sources 21 and 20, 
and 19 and 16 (at 1.4 GHz). They have not been considered to be different
components of the same source mainly because of the large ratio 
between the fluxes of the two components (sources 16 and 19) or of their
very flat spectral indices (sources 20 and 21), not consistent with the typical
spectral indices of lobes in double radio sources.
%
%{\it To allow the reader to verify the reliability of our classification,} 
%
Figure 2 shows contour maps for the 6 sources classified as double or multiple
in our samples. 

In panel (a) 1.4 and 2.4 GHz maps are shown for two sources.
Source 14(1.4 GHz)--9(2.4 GHz) presents the largest difference between the
positions found for its components at 1.4 and 2.4 GHz. In particular, the
position of component B at 1.4 GHz is displaced by $\sim$4 arcsec to the north
with respect to that
at 2.4 GHz. This may be due to the fact that at higher frequency the
emission peak could be associated to a flat spectrum "hot spot", while at lower
frequency
it is associated to the center of the lobe. Source 38(1.4 GHz)--30(2.4 GHz)
is a classical triple source at lower frequency, while is a compact single
source at higher frequency since only the core is detected above 5$\sigma$. 
In panel (b) 2.4 GHz maps are shown for the other four double or multiple
sources. Source 10 is the radio source with the higher flux ($S_{1.4 GHz} =
158$ mJy) in our sample and is identified with the quasar \# 5571
of the MZZ sample. Source 22 is a multiple
source with component B probably corresponding to the nucleus, since it has
a flatter spectral index and its position coincides with that of a red
point--like optical
object. Sources 29 and 37 are classical double sources. 
%
%{\it with no obvious optical identification helping us to better understand 
%more about their nature.}

\vspace{1cm} 

\section {The Source Counts}

The complete samples of sources with $S_p \geq 5\sigma_{local}$ at 1.4 and
2.4 GHz were used for the construction of the source counts. Complex sources
or sources with multiple components were treated as a single radio source.
Every source was weighted for the reciprocal of its visibility area (fig. 1), 
that is the area over which the source could have been seen above
the adopted limit of 5$\sigma_{local}$ (Katgert et al. 1973). 
Figure 3 shows the ratio between the total and the peak
flux as a function of flux for all the single component sources
at 1.4 GHz. For most of the sources such a ratio
is approximately normally distributed around 1, as expected for unresolved 
sources. The 1$\sigma$ dispersion of the distribution is $\sim \pm$10\% for
$S \gsimeq 1$ mJy and increases slightly at lower fluxes. The band indicated
by the dashed curves in figure 3 contains the sources which we have considered
to be unresolved. For these sources we have adopted the peak flux in
computing the source counts, while for all the others, i.e. single
component sources with $S_t/S_p$ lying above the band drawn in figure 3, and
sources clearly resolved into two or more components,
we have adopted the total flux.

In Tables 3a and 3b the resulting 1.4 GHz and 2.4 GHz source 
counts are presented. The columns give the adopted flux density intervals, 
the average flux density in each interval,
computed as the geometric mean of the two flux limits, the observed number of
sources in each flux interval, the differential source density 
(in sr$^{-1}$ Jy$^{-1}$), the normalized differential counts $nS^{2.5}$ 
(in sr$^{-1}$ Jy$^{1.5}$) with estimated errors 
(as $n^{1/2}S^{2.5}$) and the integral counts (in sr$^{-1}$). In each table
we do not report the data for the flux intervals (one at 1.4 GHz and two at
2.4 GHz) which contain only one source.

The 1.4 and 2.4 GHz normalized differential counts are plotted in figure 4a and
4b. In panel a) the solid curve represents the global fit to the counts
obtained by Windhorst, Mathis \& Neuschaufer (1990)
by fitting the counts from several 1.4 GHz surveys,
while in panel b) 
our counts are compared with the 2.7 GHz counts derived by Condon (1984), on the
basis of data by Wall, Pearson \& Longair
(1981) and with a prediction based on fluctuation analysis down to $\sim$2 mJy
obtained by Wall \& Cooke (1975). Because of the slightly different
frequencies our 2.4 GHz counts have been statistically transformed to 2.7 GHz
with an effective spectral index $\alpha = +0.53$, equal to the median value
found for the 2.4 GHz selected sample (see Section 4.2).
Note that at this frequency our counts are the deepest available, 
since the flux density limit of Wall et al. (1981) is $\sim$0.1 Jy
and the curve from Condon (1984) 
is only an extrapolation at low flux densities
based on steep spectrum and flat spectrum sources evolution models from
Wall, Pearson \& Longair (1981).

Our counts at both frequencies are in good agreement with the previous data.
%
%{\it In particular, the counts at 1.4 GHz
%flatten below about one mJy, in agreement with the well--known upturn at low
%flux density found by several authors (Windhorst 1984; Condon \& Mitchell
%1984; Windhorst et al. 1985) and clearly shown by the solid line in figure 6a.
%Such a change in slope is not visible in our 2.4 GHz counts. However,} 
%
By transforming our 2.4 GHz data to 1.4 GHz with an effective spectral index 
equal to the median value found for the 2.4 GHz selected sample,
we find that also the source counts at this frequency are consistent with the 
curve from Windhorst, Mathis \& Neuschaufer (1990), although they are not
sufficiently deep to show the well--known flattening observed in the
counts at frequencies both above and below 2.4 GHz (see Wall 1994 for a recent
discussion of counts at several frequencies). 

%
%{\it Probably a lower limiting flux at 2.4 GHz would
%be necessary in order to detect the sub--mJy flattening of the counts  
%also at this frequency. Note that also Donnelly, Partridge \& Windhorst
%(1987) do not have any clear indication of a flattening in their
%5 GHz sample with $S \gsimeq 0.1$ mJy.}
%
A Maximum Likelihood fit to our 1.4 GHz counts with two power laws:

\[ \frac{dN}{dS} \propto \left\{ \begin{array}{ll}
                   S^{-\alpha_1} & \mbox{if $S>S_b$} \\
                   S^{-\alpha_2} & \mbox{if $S<S_b$}
                   \end{array}
             \right. \]

gives the following best fit parameters: $\alpha_1 = 1.60 \pm 0.15$,
$\alpha_2 = 2.15 \pm 0.40$, $S_b \sim$0.9 mJy. Although the errors, which
represent the projection on the $\alpha_1$ and $\alpha_2$ axes of the 
1$\sigma$ combined errors on the two slopes, are relatively large,
our best fit parameters suggest that the
re--steepening of the integral counts toward an Euclidean slope starts just
below $\sim$1 mJy, in agreement with Windhorst, van Heerde \& Katgert (1984) 
and Condon \& Mitchell (1984),
while Windhorst et al. (1985, 1990), by fitting the counts to several
1.4 GHz survey, found that the change in slope starts around 5 mJy. 

%{\it As already mentioned in the Introduction, the most accepted
%interpretation of the observed flattening of the differential source counts
%is that a new population of radio sources,
%mainly faint blue galaxies often showing peculiar optical morphology, 
%appears below a few mJy. 
%The optical identifications, both photometric and spectroscopic,
%of our sub-mJy sample will be presented in a following work, which is
%presently in progress.}
%

\vspace{1cm} 

\section {The Spectral Index Distribution}

As noted before (section 2.4), 20 of the 63 sources in the 1.4 GHz sample 
were not detected at 2.4 GHz at the 5$\sigma_{local}$ level and 5
of the 48 sources in the 2.4 GHz sample were not detected at 1.4 GHz at the same
level. For these sources we inspected our maps for detection at lower
statistical level, and actually about half of them were detected at the
3$\sigma_{local}$ level. For these sources we used
their integrated flux to compute the spectral index and adopted the local
rms noise as the flux density error; for the others, which
remained undetected even at the 3$\sigma_{local}$ level, we computed
a limit on the spectral index, by assuming that their 
flux density was below the 3$\sigma_{local}$ threshold level.

Since the maps at the two frequencies have two different beams, the
fluxes given in Tables 1 and 2 do not allow a direct flux density comparison.
Therefore, we computed the spectral indices ($\alpha$; $S \propto 
\nu^{-\alpha}$) for each source by using, for the 2.4 GHz sources, the
integrated flux density obtained after convolving the 2.4 GHz image with the
same beam width as the 1.4 GHz image ($14^{\prime \prime}.5 \times 
9^{\prime \prime}.5$). For each spectral index we computed the corresponding
error as :

\begin{center}
\[ \sigma_{\alpha} = \frac{\sqrt{(\sigma_{S_1}/S_1)^2 +
(\sigma_{S_2}/S_2)^2}}{ln(\nu_1) - ln(\nu_2)}. \]
\end{center}

The data on the spectral indices are given in Table 4 for all the sources
in the two complete samples.
{\it Columns (1) and (2)} give the source numbers %instead of names (GZ). O.K.?
in the 1.4 and 2.4 GHz
catalogues, respectively; if a source is not in one of the
complete samples no number %instead of name (GZ). O.K.?
is reported at that frequency.
{\it Columns (3) and (4)} give the 1.4 and 2.4 GHz total flux densities
and corresponding 1$\sigma$ errors (in mJy). Sources which have a
listed flux in this table and not in Tables 1 and 2 have 
$3\sigma_{local} < S_p < 5\sigma_{local}$. 
{\it Column (5)} gives the spectral index (or a limit on it) and
its 1$\sigma$ error. 

Figures 5a and 5b show the spectral indices of the sources in the
complete samples at 1.4 and 2.4 GHz, respectively, as a function
of flux. Although qualitatively similar, the two distributions show
some differences:

a) As expected, the median spectral indices ($\alpha_{med}$) 
for the 2.4 GHz sample are smaller than the 1.4 GHz ones, because the higher 
frequency favours the detection of a higher fraction of flat spectrum sources. 
For both samples $\alpha_{med}$ has been computed using the ASURV package,
which implements the methods described in Feigelson and Nelson (1985) 
and Isobe, Feigelson and Nelson (1986) to statistically analyze data with 
upper or lower limits. The resulting values for $\alpha_{med}$ are
0.81 $\pm$ 0.10  for the 1.4 GHz sample and 0.53 $\pm$ 0.08 for the 2.4 GHz
sample.
%GZ: qui ho rimesso la referenza ad ASURV perche' e' richiesto dagli autori
%The values for the two samples, computed 
%with ASURV, the Survival Analysis 
%Package which uses the routines described in Feigelson and Nelson (1985) 
%and Isobe, Feigelson and Nelson (1986) and takes 
%taking into account also the
%upper or lower limits an $\alpha$, are $\alpha_{med}=0.81 \pm 0.10$ and  
%$\alpha_{med}=0.53 \pm 0.08$.

b) If each sample is divided into two about equally populated subsamples,
we find that the values for $\alpha_{med}$ at 1.4 GHz are consistent with
being constant for the high and low flux subsamples (see the $\pm 1 \sigma$
ranges for $\alpha_{med}$ in Figure 5a). This is in agreement
with the results of Donnelly, Partridge \& Windhorst (1987), who found that 
the median spectral index for their 1.46 GHz complete sample is approximately
constant at $\sim$0.75 in the whole range 0.25 $\leq S_{1.46} \leq$ 100 mJy.
Viceversa, we have some indications, at $\sim$95\% confidence level, that 
the spectra for the 2.4 GHz sample  
flatten at lower fluxes (see Figure 5b). The values of $\alpha_{med}$ 
derived for fluxes greater than and less than 0.8 mJy at this frequency
are 0.67 $\pm$ 0.09 and0.38 $\pm$ 0.11.

c) At both frequencies $\alpha_{med}$ is larger than the average
$\alpha$ because of an asymmetry in the $\alpha$ distribution due to the
presence of a non negligible number of objects with very flat or even
inverted spectra. In particular, objects with inverted spectra, which
appear at both frequencies for S $\lsimeq$ 2 mJy, constitute 
$\sim$13\% of the total 1.4 GHz sample and $\sim$25\% of the total
2.4 GHz. For this sample this percentage increases to $\sim$40\% 
for S $<$ 0.6 mJy.
These percentages are in good agreement with those found
by Donnelly, Partridge \& Windhorst (1987) at about the same flux limit: 
sources with inverted spectrum constitute $\sim$27\% and $\sim$8\%
of their complete samples, respectively at 5 and 1.4 GHz.
The tendency of an increasing percentage of
sources with inverted spectra at even lower fluxes is suggested also
by the recent result of Hammer et al. (1995), who find that
about 50\% of their optically identified $\mu$Jy radio sources
($S_{4.86GHz} > 16$ $\mu$Jy) do have inverted radio spectra. 
Such sources in their sample are mainly identified with 
early--type galaxies at z $>$ 0.75, with a smaller percentage of blue, emission--line
galaxies at lower redshifts. On the basis of the inverted spectral indices, the authors
conclude that for both classes of objects the radio emission is powered by a mini--AGN
rather than by starburst activity. All the early--type galaxies with inverted spectrum in the
Hammer et al. sample have very faint $V$ magnitude, in the range $23 \lsimeq V \lsimeq 26$.
Similar objects in the existing sub--mJy samples would have not been spectroscopically
identified. To verify whether the objects with inverted spectrum in our sample are the
"bright" counterparts of the objects found in the $\mu$Jy sample, we are obtaining deep
spectroscopic observations at the ESO 3.6--m telescope (Gruppioni et al., in preparation).
On the basis of preliminary data for about 40\% of our complete samples 
of radio sources, we indeed find that almost all of the spectroscopically
observed sources with inverted radio spectrum are identified with early--type 
galaxies
at intermediate redshift. Enhanced starburst activity, while present in a few 
sources
with steep radio spectral index ($\alpha > 0.5$), is not seen in any of the
optical counterparts of sources with inverted radio spectrum.

\vspace{1cm}

\section {Conclusions}

In this paper we have presented the first deep survey obtained at two
frequencies (1.4 and 2.4 GHz) with the ATCA. With the use of a mosaic
technique (four separate observations with different pointing positions)
we have been able to obtain a reasonably low 5$\sigma_{local}$ limiting
flux over an area large enough ($\sim$0.36 sq. deg.) to provide acceptable
statistics at each frequency. The complete samples of 5$\sigma_{local}$
sources are constituted by 63 and 48 sources at 1.4 and 2.4 GHz respectively.

The main results of our analysis of the radio data are:

i) Our normalized differential counts at 1.4 GHz show a flattening
below about one mJy, in agreement with the results from other 1.4 GHz deep surveys
(Condon \& Mitchell 1984; Condon 1984; Windhorst 1984; Windhorst et al. 1985).
A formal Maximum Likelihood fit to our counts suggests a resteepening toward an
Euclidean
slope, although with a relatively large error ($\alpha = 2.15 \pm 0.40$),
just below $\sim$1 mJy.

ii) Our 2.4 GHz counts are the deepest at this or similar frequencies (i.e. 2.7 GHz).
They agree with previous counts at significantly higher flux (Wall, Pearson \&
Longair 1981; Condon 1984; Wall \& Peacock 1985). Although no change in slope
is seen in the 2.4 GHz counts, they are however consistent with the 
best fit curve at lower frequency.

iii) Dividing each sample into two about equally populated subsamples, we find that
the values for $\alpha_{med}$ at 1.4 GHz are consistent with being constant for the high
and low flux subsamples, in agreement with the results of Donnelly, Partridge
\& Windhorst (1987).
Viceversa, we have some indications (significant at $\sim$2$\sigma$ level) that
at 2.4 GHz $\alpha_{med}$
decreases at lower fluxes. Note that, since in analysing the distributions of the
spectral indices for the complete samples at each frequency we fully used also the
information given by the presence of a few limits on $\alpha$, these results should
not be affected by the biases discussed in Donnelly, Partridge \& Windhorst (1987).

iv) At both frequencies $\alpha_{med}$ is larger than the average $\alpha$ because
of an asymmetry in the $\alpha$ distribution due to the presence of a non negligible
number of objects with very flat or even inverted spectra. In particular, objects with
inverted spectra, which appear at both frequencies for $S \lsimeq 2$ mJy, constitute
$\sim$13\% of the total 1.4 GHz sample and $\sim$25\% of the total 2.4 GHz one.
For this sample this percentage increases to $\sim$40\% for $S < 0.6$ mJy. The tendency
of an increasing percentage of sources with inverted spectra at even lower fluxes is
confirmed by the recent result of Hammer et al. (1995), who find that about 50\% of
their optically identified $\mu$Jy radio sources ($S_{4.86GHz} > 16 \mu$Jy) do have
inverted radio spectra.

Optical identifications of our radiosources and, in particular, of our 
inverted--spectrum faint sources should help in better understanding the
nature of this 
%new GZ : qui io toglierei il "new"; vedi tu se vuoi tenerlo
population of objects. The results of such work will be
presented in a following paper (Gruppioni et al., in preparation), which is 
presently in progress.

\vspace{1cm}

\section {Acknowledgements}

This paper is based on observations collected at the Australia Telescope
Compact Array (ATCA), which is operated by the CSIRO Australia Telescope
National Facility. 
The authors thank Roberto Fanti for helpful suggestions and
for a careful reading of the manuscript. 
This work has been partially financed by the Italian Space Agency (ASI)
under the contract ASI--1995--RS--152.

\label{lastpage}
 
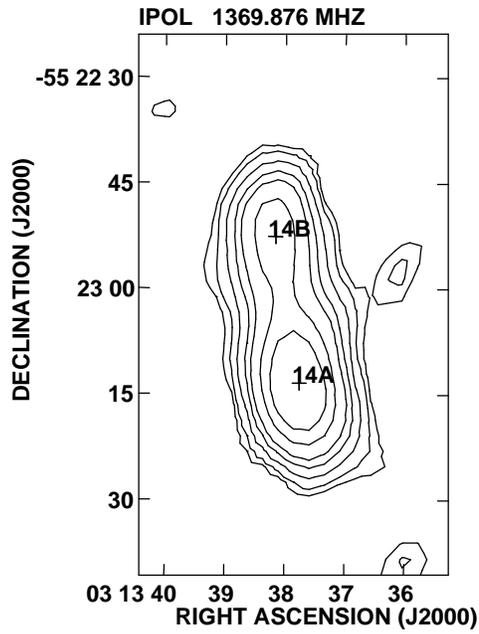
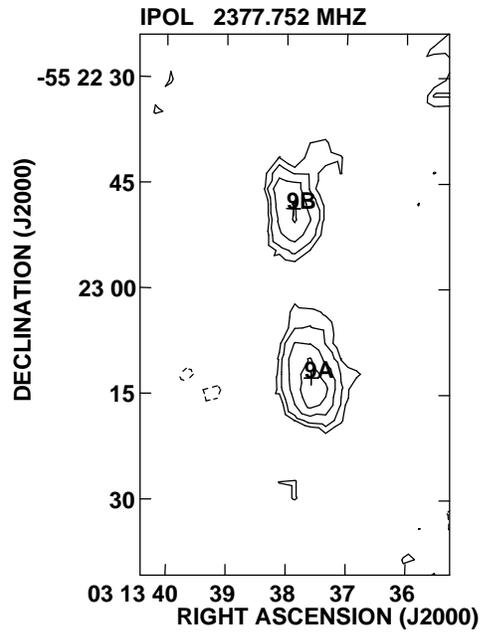
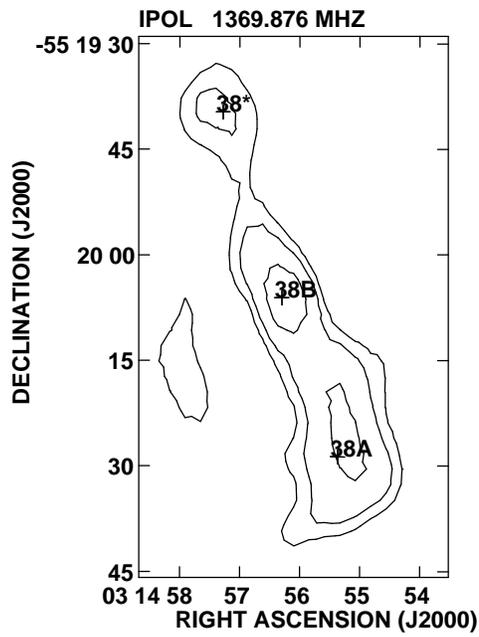
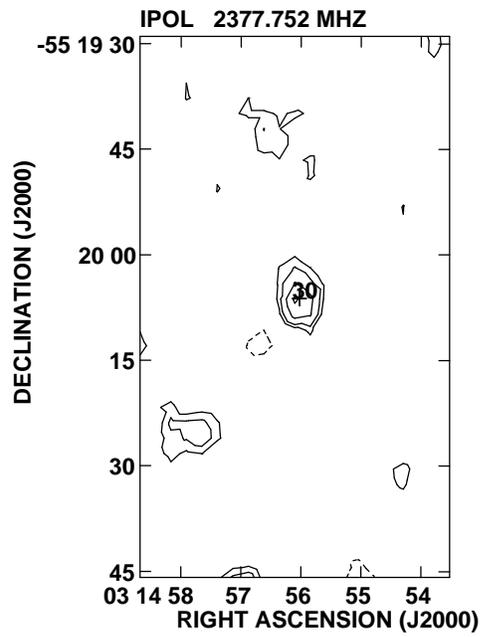
\end{document}